\documentstyle[12pt]{article}
\begin{document}

\title{Reasoning about quantum systems\\ at the macroscopic level}

\author{Jochen Rau\thanks{email: jochen.rau@q-info.org} 
}

\date{\today}

\maketitle

\begin{abstract}

In this didactical note I review in depth the rationale for using generalised canonical distributions in quantum statistics.
Particular attention is paid to the proper definitions of quantum entropy and quantum relative entropy, as well as to quantum state reconstruction on the basis of incomplete data.
There are two appendices in which I outline how generalised canonical distributions link to the conventional formulation of statistical mechanics,
and how classical probability calculus emerges at the macroscopic level.
\end{abstract}

\newpage

\section{\label{introduction}What is the problem?}

Reasoning with probabilities requires two basic algorithms:
(i) 
a rule for updating probabilities when new evidence becomes available;
and
(ii)
a prescription for determining the starting point, i.e., for constructing the initial probability distribution on the basis of---usually incomplete---prior knowledge.
In classical statistical inference\footnote{
For excellent introductions to classical statistical inference see, e.g., the very readable book by Sivia \cite{sivia:book} and the seminal work of Jaynes \cite{jaynes:papers,jaynes:book}.
}
these two ingredients are furnished by \textit{Bayes rule} and the \textit{maximum entropy principle}, respectively.
{Bayes rule} 
\begin{equation}
	\mbox{\rm prob}(\mbox{\rm hypothesis}|\mbox{\rm data})=
	\frac{\mbox{\rm prob}(\mbox{\rm data}|\mbox{\rm hypothesis})\cdot \mbox{\rm prob}(\mbox{\rm hypothesis})}{\mbox{\rm prob}(\mbox{\rm data})}
\end{equation}
stipulates how probabilities are to be updated in the light of new data, thus encapsulating the process of learning;
whereas the {maximum entropy principle} provides the starting point for this learning process by assigning to a hypothesis $i$ the prior probability
\begin{equation}
	q_i = \frac{1}{Z} \exp\left[-\sum_{a=1}^m \lambda^a G^i_a\right]
	\quad,\quad i=1\ldots d
\label{maxent:classical}
\end{equation}
with \textit{partition function} 
\begin{equation}
	Z:=\sum_{i=1}^d \exp\left[-\sum_{a=1}^m \lambda^a G^i_a\right] \quad.
\end{equation}
Such a \textit{generalised canonical distribution} maximises the classical entropy
\begin{equation}
	S[\{q_i\}]:=-\sum_{i=1}^d q_i \ln q_i
\label{entropy:classical}
\end{equation}
under the normalisation condition $\sum_i q_i=1$ and the $m$ ($m<d$) linearly independent constraints
\begin{equation}
	\langle G_a\rangle_q:= \sum_{i=1}^d q_i G^i_a = g_a\quad,\quad a=1\ldots m
\end{equation}
which are deemed the only prior information available.
These constraints uniquely specify the $m$ Lagrange parameters $\{\lambda^a\}$.

Quantum theory is a full-fledged probabilistic theory that despite its counterintuitive features shares with classical probability calculus a high degree of internal consistency \cite{caves:quantumasbayes,rau:qvc}.
It should therefore be possible to erect the edifice of quantum statistical inference on the same two pillars.
Indeed:

\begin{enumerate}

\item
There is a quantum analog of the classical Bayes rule \cite{schack:bayesrule}.
This ``quantum Bayes rule'' pertains to experiments on exchangeable sequences.
An exchangeable sequence of length $N$ can be thought of informally as a finite subsequence of an infinite sequence of systems whose order is irrelevant.
It has a probability distribution of the de Finetti form \cite{caves:definettistates}
\begin{equation}
	\rho^{(N)}=\int_{{\cal S}(d)}d\rho\:{\rm prob}(\rho)\:\rho^{\otimes N}\ ,
\label{definetti}
\end{equation}
where the ``meta-probability'' ${\rm prob}(\rho)\geq 0$ is normalised to
\begin{equation}
	\int_{{\cal S}(d)}d\rho\:{\rm prob}(\rho)=1\ 
\end{equation}
and the integration is over the manifold ${\cal S}(d)$ of probability distributions.
After ascertaining the outcome $\Gamma^{(K)}$ of $K$ ($K<N$) trials (i.e., of some measurement performed on $K$ constituents of the sequence) the posterior $\rho^{(N-K)}$ for the remaining $(N-K)$ constituents\footnote{
not for all $N$ constituents, because $K$ constituents have been disturbed by quantum measurement
}
still has the de Finetti form, yet with a new meta-probability that has been updated according to the quantum Bayes rule
\begin{equation}
	{\rm prob}(\rho|\Gamma^{(K)}) =
	\frac{\mbox{\rm prob}(\Gamma^{(K)} | \rho) \cdot {\rm prob}(\rho)}{\mbox{\rm prob}(\Gamma^{(K)})}
	\quad,
\label{metabayes}
\end{equation}
where $\mbox{\rm prob}(\Gamma^{(K)} | \rho) := \mbox{\rm prob}(\Gamma^{(K)} | \rho^{\otimes K})$ and
\begin{equation}
	\mbox{\rm prob}(\Gamma^{(K)}) :=
	\int_{{\cal S}(d)}d\rho'\;\mbox{\rm prob}(\Gamma^{(K)} | \rho') \cdot \mbox{\rm prob}(\rho')
	\quad.
\label{old17}	
\end{equation}

\item
There is a quantum version of the maximum entropy prior, namely the generalised canonical statistical operator \cite{balian:book1}
\begin{equation}
	{\rho}=\frac{1}{Z} \exp\left[-\sum_{a=1}^m \lambda^a {G_a}\right]
\label{maxent:quantum}
\end{equation}
with
\begin{equation}
	Z:=\mbox{\rm tr}\left\{\exp\left[-\sum_{a=1}^m \lambda^a {G_a}\right]\right\} \quad,
\end{equation}
which maximises the \textit{von Neumann entropy}
\begin{equation}
	S[{\rho}]:=-\mbox{\rm tr} ({\rho}\ln{\rho})
\label{old5}
\end{equation}
under the constraints $\mbox{\rm tr}\rho=1$ and
\begin{equation}
	\langle G_a\rangle_\rho:= \mbox{\rm tr} ({\rho}{G_a})=g_a\quad,\quad a=1\ldots m\quad.
\label{constraints:quantum}
\end{equation}

\end{enumerate}
\noindent
It is the foundations of the second pillar that shall be inspected more closely in this didactical note.

Strictly speaking, in both the classical and the quantum case the use of the generalised canonical form (\ref{maxent:classical}) or (\ref{maxent:quantum}) rests on two tacit assumptions which are often, but not always justified:
(i)
that the probability distribution is indeed normalised, $\sum_i q_i=1$ or $\mbox{\rm tr}\rho=1$;
and
(ii)
that ``total ignorance'' (i.e., the absence of any constraints, $\lambda^a=0$) must correspond to a uniform distribution $q_i=1/d$ or $\rho=1/d$, with $d:=\mbox{\rm tr} I$ in the latter case.
When these two assumptions are relaxed then the prior must have the more general form (in the quantum case)
\begin{equation}
			\rho =\frac{\iota}{Z} \exp\left[\left(\ln\sigma-\langle\ln\sigma\rangle_{1/d}\right) - \sum_{a=1}^m \lambda^a G_a\right]
\label{old159}
\end{equation}
with partition function
\begin{equation}
	Z:=\mbox{\rm tr}\left\{
	\exp\left[\left(\ln\sigma-\langle\ln\sigma\rangle_{1/d}\right) -\sum_{a=1}^m \lambda^a G_a\right] \right\}
	\quad,
\end{equation}
where the state $\sigma$ represents ``total ignorance'', i.e., the---possibly non-uniform---starting distribution in the absence of any constraints ($\lambda^a=0$).
This more general prior, rather than maximising the von Neumann entropy (\ref{old5}), minimises the \textit{quantum relative entropy}\footnote{
The following notation is not universal but the most commonly used in the modern literature on quantum information theory \cite{nielsen:book}.
Some authors, e.g. \cite{peres:book}, also use $S(\sigma | \rho)\equiv S(\rho\|\sigma)$.
}
\begin{equation}
	S(\rho\|\sigma) :=
	\mbox{\rm tr}({\rho} \ln {\rho} - {\rho} \ln {\sigma}) 
\label{old8}
\end{equation}
with respect to $\sigma$ under the constraints (\ref{constraints:quantum}) and $\mbox{\rm tr}\rho=\iota\in(0,1]$ \cite{ruskai:minrent};
it reduces to the familiar canonical form (\ref{maxent:quantum}) whenever $\sigma=1/d$ and $\iota=1$.
The generalisation of the classical case is completely analogous and involves minimising the classical relative entropy
\begin{equation}
	S(\{q_i\}\|\{p_i\}):= \sum_{i=1}^d q_i \ln\frac{q_i}{p_i}\quad.
\label{2star}
\end{equation}
The need to consider such more general situations is particularly apparent in the case of classical \textit{continuous} distributions, 
\begin{equation}
	q_i\to \pi(x)\quad,\quad \sum_i\to \int dx\quad,
\end{equation}
where---depending on the coordinates chosen---``total ignorance'' need no longer correspond to a uniform distribution.
The extremisation of relative rather than ordinary entropy, and hence the use of minimum relative entropy (MinREnt) rather than maximum entropy (MaxEnt) priors then ensures that any conclusions drawn from statistical inference are coordinate-independent \cite{rau:physrep}.

In most textbooks on statistical mechanics the derivation of the quantum state (\ref{maxent:quantum})---or of the more general state (\ref{old159})---proceeds heuristically by simple analogy with the classical case.
Yet at closer inspection the justification for using these states is far from obvious.
Specifically, the quantum case presents two major, interrelated difficulties:

\begin{itemize}

\item
The von Neumann entropy (\ref{old5}) is not the only possible ``quantisation'' of the classical entropy (\ref{entropy:classical}) and hence not necessarily the quantity to be maximised;
there are various other conceivable definitions of quantum entropy that reduce to Eq. (\ref{entropy:classical}) in the classical limit.
Likewise, in the more general setting involving non-uniform ignorance distributions $\sigma$ there are various conceivable definitions of quantum relative entropy that all reduce to Eq. (\ref{2star}) in the classical limit, and that hence are candidates for minimisation.
Singling out the definitions (\ref{old5}) and (\ref{old8}) as the correct quantum analogs of Eqs. (\ref{entropy:classical}) and (\ref{2star}) is a non-trivial task \cite{ochs:axiomatic,donald:cmp,hiai+petz}.

\item
The use of the quantum state (\ref{maxent:quantum}) or (\ref{old159}), respectively, implies the assertion that such a generalised canonical distribution is indeed most typical of the states allowed by the constraints (\ref{constraints:quantum}); 
i.e., that whenever the expectation values (\ref{constraints:quantum}) have been ascertained as measured averages in sufficiently many trials on an exchangeable sequence, the posterior single-constituent state must be close to this generalised canonical state.
While in the classical case this can be understood with the help of Jaynes' ``entropy concentration theorem'' \cite{jaynes:concentration,rau:nutshell} or the didactical ``monkey'' and ``kangaroo'' arguments \cite{sivia:book},\footnote{
The logic of these statistical arguments is similar to the textbook treatment of the classical canonical distribution as describing just one constituent of some larger microcanonical ensemble.
}
the extension to the quantum case---especially if the constraints pertain to non-commuting observables---is far from straightforward \cite{balian+balazs,buzek:reconstruction}.

\end{itemize}
\noindent
In the following two sections I will address these issues---the proper quantisation of entropies and the alleged typicality of canonical quantum states---one by one.
There will also be two short appendices linking these fundamental concepts to the conventional formulation of statistical mechanics
and sketching the emergence of classicality in the macroscopic limit, respectively.

\section{\label{quantumentropies}Quantum entropies}

The concepts of entropy and relative entropy play a pivotal role in mathematical physics, both in statistical mechanics \cite{balian:book1} and in modern quantum information theory \cite{nielsen:book,schumacher+westmoreland,vedral:entanglement,vedral:rmp}.
Since the ordinary entropy $S[\rho]$ can be expressed in terms of the relative entropy, 
\begin{equation}
	S[\rho]=S[1/d]-S(\rho\|1/d)
\label{old75}
\end{equation}
---with calibration such that $S[|\psi\rangle\langle\psi|]=0$ for arbitrary pure states $\psi$---,
the problem of quantising the former can be reduced to that of quantising the latter.
Finding the proper formula for quantum relative entropy, in turn, can be approached from two different angles which I shall discuss separately in the following subsections.
The first approach makes use of a central result of quantum information theory, the so-called quantum Stein lemma.
The second approach is axiomatic in nature and starts from a set of consistency requirements for meta-probabilities.
There is a third subsection in which I list some useful properties of quantum entropies.

\subsection{\label{lemma}Quantum Stein lemma}

I start out by recapitulating how the concept of relative entropy emerges in classical statistical inference, following closely my earlier didactical note \cite{rau:nutshell}.
Given a classical prior probability distribution $\{p_i\}$ for the results $i=1\ldots d$,
the probability that $N$ trials will yield the---generally different---relative frequencies $\{f_i=N_i/N\}$ is
\begin{equation}
\mbox{prob}(\{f_i\}|\{p_i\},N)
=
{N!\over N_1!\ldots N_{d}!}\,
p_1^{N_1}\ldots p_{d}^{N_{d}}
\quad.
\end{equation}
Here the second factor is the probability for one specific outcome with sample numbers $\{N_i\}$,
whereas the first factor counts the number of all outcomes that give rise to the same set of sample numbers. 
With the definition (\ref{2star}) of classical relative entropy
and the shorthand notations $f\equiv\{f_i\}$, $p\equiv\{p_i\}$, as well as $f^{\otimes N}$, $p^{\otimes N}$ for composite distributions pertaining to $N$ trials, one can also write
\begin{equation}
\mbox{prob}(f^{\otimes N}|p^{\otimes N})
=
\mbox{prob}(f^{\otimes N}|f^{\otimes N})\exp[-N S(f\|p)]
\quad.
\label{ways}
\end{equation}
By virtue of Stirling's formula
\begin{equation}
x!\approx \sqrt{2\pi x}\, x^x e^{-x}
\end{equation}
the pre-factor can be approximated by
\begin{equation}
	\mbox{prob}(f^{\otimes N}|f^{\otimes N})\approx (2\pi N)^{-(d-1)/2} \prod_{i=1}^{d} \sqrt{\frac{1}{f_i}}
	\sim O\left(N^{-(d-1)/2}\right)
\label{old22}
\end{equation}
and thus scales as $N^{-({d}-1)/2}$ with the number of trials.
For $N\to\infty$ the meta-distribution $\mbox{prob}(f^{\otimes N}|p^{\otimes N})$ of frequencies becomes completely dominated by the exponential $\exp[-N S(f\|p)]$.
Then the probability with which any given frequency distribution $f$ is realized is essentially determined by the quantity $S(f\|p)$:
The larger this quantity, the less likely the frequency distribution is realized.
Since $S(f\|p)\geq 0$ with equality if and only if $f=p$, the meta-distribution becomes sharply peaked around $f=p$.

The observed relative frequencies $\{f_i\}$ may be visualized as Cartesian coordinates of a point in a $d$-dimensional vector space,
where $f_i\in[0,1]$ and the normalization condition $\sum_i f_i=1$ restrict the allowed points to some portion of a $(d-1)$-dimensional hyperplane.
In this hyperplane portion there is a unique point, namely $f=p$, at which the quantity $S(f\|p)$ vanishes;
everywhere else $S(f\|p)$ is strictly positive.
It is possible to define new coordinates $\{x_1\ldots x_{d-1}\}$ in the hyperplane such that
(i)
they are linear functions of the $\{f_i\}$;
(ii)
the origin ($\vec{x}=0$) is at $f=p$; and
(iii)
in the vicinity of $f=p$,
\begin{equation}
S(f(\vec{x})\|p)=a r^2 + O(r^3)
\quad,\quad
a>0
\quad,
\end{equation}
where
\begin{equation}
r:=\sqrt{\sum_{j=1}^{d-1} x_j^2}
\quad.
\end{equation}
Frequency distributions whose $S(f(\vec{x})\|p)$ exceeds some finite threshold $\Delta\! S$ thus lie outside a hypersphere around $f=p$,
the sphere's radius $R$ being given by $aR^2=\Delta S$.
The probability that $N$ trials will yield such a frequency distribution outside the hypersphere is 
\begin{equation}
\mbox{prob}[S(f\|p)>\Delta\! S|(d-1),N]
=
\frac{\int_R^\infty dr\,r^{d-2}\exp(-Nar^2)}{\int_0^\infty dr'\,{r'}^{d-2}\exp(-Na{r'}^2)}
\quad.
\end{equation}
Here $(d-1)$ is noted as the dimension of the hyperplane.
The factors $r^{d-2}$ in the integrand are due to the volume element, while the exponentials
$\exp(-Nar^2)$ 
stem from the asymptotically dominant exponential factor in the meta-probability (\ref{ways}).
Substituting $t:=Nar^2$ and using
\begin{equation}
\Gamma\left[\frac{d-1}{2}\right]=\int_0^{\infty} dt\,t^{\frac{d-1}{2}-1}\exp(-t)
\end{equation}
one may also write
\begin{equation}
\mbox{prob}[S(f\|p)>\Delta\! S|(d-1),N]
= \frac{1}{\Gamma\left[\frac{d-1}{2}\right]} {\int_{N\Delta S}^\infty dt\,t^{\frac{d-1}{2}-1}\exp(-t)}
\quad;
\end{equation}
which for large $N$ ($N\gg d/\Delta S$) can be approximated
by
\begin{equation}
\mbox{prob}[S(f\|p)>\Delta\! S|(d-1),N]
\sim
\frac{1}{\Gamma\left[\frac{d-1}{2}\right]} (N\Delta S)^{\frac{d-1}{2}-1} \exp(-N\Delta S)
\, .
\label{approx}
\end{equation}
As the number $N$ of trials increases, this probability rapidly tends to zero for any finite $\Delta S$. 
As $N\to\infty$, therefore, it becomes virtually certain that the measured frequency distribution $f$ has $S(f\|p)$ very close to zero, and hence coincides with the prior $p$.
So not only does $f=p$ represent the frequency distribution that is the most likely to be realized (cf. Eq. (\ref{ways}));
but in addition, as $N$ increases, all other---theoretically allowed---frequency distributions become more and more concentrated near $f=p$:
Frequency distributions other than $f=p$ become highly atypical.
Entropy fluctuations around $f=p$ have decreasing size 
\begin{equation}
	\langle S(f\|p)\rangle_f =
	\frac{\int_0^\infty dr\,r^{d-2}\,ar^2\exp(-Nar^2)}{\int_0^\infty dr'\,{r'}^{d-2}\exp(-Na{r'}^2)}
	= \frac{d-1}{2}\cdot \frac{1}{N}
\label{old29}	
\end{equation}
of order $O(1/N)$, which due to their quadratic dependence on $|f-p|$ correspond to a frequency range $|f-p|\sim O(1/\sqrt{N})$.
These results are known as the ``entropy concentration theorem'' \cite{jaynes:concentration}.

In the asymptotic regime $N\to\infty$ one may formulate the following hypothesis
$\Gamma_q^N$: ``After $N$ trials the measured relative frequencies are within $O(1/\sqrt{N})$ around $q$.''
This frequency range corresponds to relative entropies up to $\Delta S\sim O(1/N)$ with respect to $q$.
According to the entropy concentration theorem the hypothesis is almost certainly true in the state $q^{\otimes N}$,
\begin{equation}
	\mbox{\rm prob}(\Gamma_q^N|q^{\otimes N})\sim 1-\mbox{prob}[S(f\|q)>1/N|(d-1),N] \sim O(1)
	\quad,
\end{equation}
independent of $N$.
Let $p\neq q$ be some other distribution that is a finite distance away from $q$, $|p-q|\sim O(1)$.
In this different state the probability becomes 
\begin{eqnarray}
	\mbox{\rm prob}(\Gamma_q^N|p^{\otimes N})&=& \sum_{f\in\Gamma_q^N} \mbox{prob}(f^{\otimes N}|p^{\otimes N}) \nonumber\\
	&=& \sum_{f\in\Gamma_q^N} \mbox{prob}(f^{\otimes N}|f^{\otimes N}) \exp[-N S(f\|p)]
	\quad.
\end{eqnarray}
Since measurable relative frequencies are spaced at intervals of size $O(1/N)$,
the summation is over $O([(1/\sqrt{N})/(1/N)]^{(d-1)})\sim O(N^{(d-1)/2})$ different measurable distributions;
which combined with the asymptotics of Eq. (\ref{old22}) implies
\begin{equation}
	\sum_{f\in\Gamma_q^N} \mbox{prob}(f^{\otimes N}|f^{\otimes N}) \sim O(1)\quad.
\end{equation}
Thus the asymptotic dependence on $N$ is determined entirely by the exponential function.
Its argument may be expanded around $f=q$ in powers of $(f-q)\sim O(1/\sqrt{N})$ and for $N\to\infty$ is dominated by the leading term $[-NS(q\|p)]$.
In this limit it is therefore
\begin{equation}
	\lim_{N\to\infty} \frac{1}{N}\ln\,\mbox{\rm prob}(\Gamma_q^N|p^{\otimes N}) = -S(q\|p)
	\quad.
\end{equation}

Due to the entropy concentration theorem any hypothesis $\Gamma^N$ which is true in $q^{\otimes N}$ in the asymptotic sense, $\mbox{\rm prob}(\Gamma^N|q^{\otimes N})\sim O(1)$, must relate to a frequency range that encloses the $O(1/\sqrt{N})$-neighborhood of $q$.
Hence $\Gamma^N\supseteq \Gamma^N_q$, and $\mbox{\rm prob}(\Gamma^N|p^{\otimes N})\geq \mbox{\rm prob}(\Gamma_q^N|p^{\otimes N})$ for any $p$.
Consequently the above limit can also be expressed as
\begin{equation}
	\lim_{N\to\infty} \frac{1}{N}\ln\,\inf \left\{\mbox{\rm prob}(\Gamma^N|p^{\otimes N}) \right|\left. \mbox{\rm prob}(\Gamma^N|q^{\otimes N})\sim O(1)\right\} 
	= -S(q\|p)
	\quad.
\label{classical:stein}
\end{equation}
It is this result which allows one to build a bridge to the quantum case.
In close analogy to the above classical case one can consider the set of hypotheses $\Gamma^N$---now represented by projection operators or, more generally, positive operators---that pertain to $N$-partite sequences, and whose probabilities in some state $\rho^{\otimes N}$ are of order $O(1)$, i.e., always larger than some finite threshold $1-\epsilon$ ($0<\epsilon<1$) regardless of the number $N$ of trials.
The probability that such a hypothesis is found true in a \textit{different} state $\sigma^{\otimes N}$ has a lower bound that satisfies
\begin{equation}
	\lim_{N\to\infty} \frac{1}{N}\ln\,\inf \left\{\mbox{\rm tr}(\sigma^{\otimes N}\Gamma^N)\right|\left. 0\leq\Gamma^N\leq I,\, \mbox{\rm tr} (\rho^{\otimes N}\Gamma^N)\geq 1-\epsilon \right\}
	= -S(\rho\|\sigma)
\label{quantum:stein}
\end{equation}
independently of the precise value of $\epsilon$, where $S(\rho\|\sigma)$ is the quantum relative entropy as defined in Eq. (\ref{old8}).
This is the ``quantum Stein lemma''.
Its proof is rather intricate due to the possibility that the observables used to prepare the prior and the observables being  subsequently measured need not commute, and can be found elsewhere \cite{hiai+petz,ogawa+nagaoka,hayashi:book,petz:book}.

The limits (\ref{classical:stein}) and (\ref{quantum:stein}) suggest a common interpretation of relative entropy in both the classical and quantum cases.
The infimum on the left-hand side may be regarded, loosely, as the probability that despite prior preparation of a system in the state $p$ or $\sigma$, a high-resolution measurement on $N$ replicas will yield an outcome that corresponds to a \textit{different} state (or range of states within an $O(1/\sqrt{N})$-neighborhood of) $q$ or $\rho$, respectively.
Asymptotically, with each additional trial the probability for such deviating evidence decreases further by a factor $\exp[-S(q\|p)]$ or
\begin{equation}
	\mbox{\rm prob}(\rho|\sigma):=\exp[-S(\rho\|\sigma)]\quad,
\label{3star}	
\end{equation}
respectively.

\subsection{Axiomatic approach}

The axiomatic approach to quantisation takes the interpretation (\ref{3star}) not as a result but as its starting point.
This approach formulates a number of consistency requirements either for the quantum relative entropy directly \cite{donald:cmp} or---as I shall prefer to do here---for the meta-probability $\mbox{\rm prob}(\rho|\sigma)$.
These consistency requirements lead unequivocally to the proper definition of quantum relative entropy.

The following four requirements for meta-probabilities are straightforward:

\begin{enumerate}

\item
The probability for evidence $\rho$ vanishes whenever it lies outside the support of the prior;
or in reverse,
\begin{equation}
	\mbox{\rm prob}(\rho|\sigma)>0\ \Leftrightarrow\ \mbox{\rm supp}\;\rho\subseteq\mbox{\rm supp}\;\sigma \quad.
\label{probability:support}
\end{equation}

\item
Meta-probabilities are invariant under joint unitary transformations $g$,
\begin{equation}
	\mbox{\rm prob}(g(\rho)|g(\sigma))=\mbox{\rm prob}(\rho|\sigma)
	\quad.
\label{groupinvariant}
\end{equation}

\item
Provided there is a hypothesis $a$ such that both $\mbox{\rm supp}\;\rho\subseteq a$ and $\mbox{\rm supp}\;\sigma\subseteq a$, meta-probabilities do not change when the Hilbert space is reduced and distributions are restricted to $a$:
\begin{equation}
	\mbox{\rm prob}\left(\rho|_a\;\right|\left. \sigma|_a\right)=\mbox{\rm prob}(\rho|\sigma)
	\quad.
\label{embedding}
\end{equation}

\item
For uncorrelated prior and posterior distributions the meta-probability factorises,
\begin{equation}
	\mbox{\rm prob}(\rho_A\otimes\rho_B|\sigma_A\otimes\sigma_B)=\mbox{\rm prob}(\rho_A|\sigma_A)\cdot \mbox{\rm prob}(\rho_B|\sigma_B)
	\quad.
\label{productrule}
\end{equation}

\end{enumerate}

\noindent
Two additional requirements are less obvious and require further motivation:

\begin{enumerate}
\setcounter{enumi}{4}

\item
The degrees of freedom that completely specify a probability distribution can often be divided into some that are actually being prepared, measured or otherwise considered ``relevant'', and the rest deemed ``irrelevant''.
By eliminating all information pertaining to the irrelevant degrees of freedom an arbitrary probability distribution can be reduced to its \textit{relevant part}, a procedure known as \textit{coarse-graining}.
Two examples for such coarse-graining are
\begin{enumerate}
\renewcommand{\theenumii}{\roman{enumii}}
\item
discarding all information except for some selected probabilities $\{\mbox{\rm tr}(\rho P_i)\}$, where the $\{P_i\}$ are projectors onto mutually orthogonal, collectively exhaustive (i.e., $\sum_i P_i=I$) subspaces.
The relevant part of a statistical operator $\rho$ is then
\begin{equation}
	{\cal P}_{\{P_i\}} \rho:=\sum_i \frac{\mbox{\rm tr}(\rho P_i)}{\mbox{\rm tr} P_i} P_i
	\quad;
\label{4star}	
\end{equation}
\item
removing from the state $\rho_{AB}$ of a composite system all correlations between the subsystems $A$ and $B$, yielding the relevant part
\begin{equation}
	{\cal P}_{\{X\otimes I\}} \rho:= \frac{(\mbox{\rm tr}_B\rho_{AB})\otimes(\mbox{\rm tr}_A\rho_{AB})}{\mbox{\rm tr}_{AB}\rho_{AB}}
	\quad.
\label{4startilde}	
\end{equation}
\end{enumerate}
For a generic coarse-graining I shall write $\rho\to{\cal P}\rho$.
The map ${\cal P}$, acting on the manifold of probability distributions, must satisfy ${\cal P}^2={\cal P}$, leave the expectation values of relevant observables unchanged, and under these constraints maximise the resemblance of ${\cal P}\rho$ with the (often, but not always uniform) ``ignorance distribution'', i.e., maximise the likelihood $\mbox{\rm prob}({\cal P}\rho|\mbox{\rm ignorance})$.
Specifying the relevant part ${\cal P}\rho$ is thus tantamount to specifying a system's relevant degrees of freedom.
When the latter are prepared in some state ${\cal P}\sigma$, the probability that subsequent experimental evidence will correspond to a full state $\rho$, and that data for the relevant degrees of freedom will correspond to a relevant part ${\cal P}\rho$, is according to Bayes rule
\begin{equation}
	\mbox{\rm prob}(\rho,{\cal P}\rho|{\cal P}\sigma)= \mbox{\rm prob}(\rho|{\cal P}\rho,{\cal P}\sigma)\cdot \mbox{\rm prob}({\cal P}\rho|{\cal P}\sigma)
	\quad.
\end{equation}
Yet on the left-hand side ${\cal P}\rho$ is redundant because it is implied by $\rho$;
and in the first factor on the right-hand side ${\cal P}\sigma$ is redundant because it is superseded by the subsequent deviating evidence ${\cal P}\rho$.
These considerations motivate the ``chain rule''
\begin{equation}
	\mbox{\rm prob}(\rho|{\cal P}\sigma)= \mbox{\rm prob}(\rho|{\cal P}\rho)\cdot \mbox{\rm prob}({\cal P}\rho|{\cal P}\sigma)
	\quad.
\end{equation}

\item
For the particular coarse-graining (\ref{4star}) the meta-probability $\mbox{\rm prob}({\cal P}\rho|{\cal P}\sigma)$ does not depend on the precise orientation or dimensionality of the subspaces associated with $\{P_i\}$ but only on the respective sets of relevant probabilities $\{\mbox{\rm tr}(\rho P_i)\}$ and $\{\mbox{\rm tr}(\sigma P_i)\}$; 
so
\begin{equation}
	\mbox{\rm prob}({\cal P}_{\{P_i\}}\rho|{\cal P}_{\{P_i\}}\sigma)=
	\mbox{\rm prob}(\{\mbox{\rm tr}(\rho P_i)\}|\{\mbox{\rm tr}(\sigma P_i)\})
	\quad,
\end{equation}
where the right-hand side is the classical counterpart of definition (\ref{3star}).

\end{enumerate}

By virtue of definition (\ref{3star}) the above six requirements for meta-probabilities translate into properties of the quantum relative entropy:

\begin{enumerate}

\item
\textit{Range:}
\begin{equation}
	S(\rho\|\sigma)\left\{
\begin{array}{ll}
	\in [0,\infty) & : \mbox{\rm supp}\;\rho\subseteq\mbox{\rm supp}\;\sigma \\
	= +\infty & : \mbox{\rm otherwise}
\end{array}
\right.
\quad.
\label{positivity:rent}
\end{equation}

\item
\textit{Unitary invariance:}
\begin{equation}
	S(g(\rho)\|g(\sigma))=S(\rho\|\sigma)
	\quad.
\label{groupinvariant:rent}
\end{equation}

\item
\textit{Invariance under Hilbert space reduction:}
If both $\mbox{\rm supp}\;\rho\subseteq a$ and $\mbox{\rm supp}\;\sigma\subseteq a$ then
\begin{equation}
	S\left(\rho|_a\right\|\left.\sigma|_a\right)=S(\rho\|\sigma)
	\quad.
\label{embedding:rent}
\end{equation}

\item
\textit{Additivity:}
\begin{equation}
	S(\rho_A\otimes\rho_B\|\sigma_A\otimes\sigma_B)=S(\rho_A\|\sigma_A)+S(\rho_B\|\sigma_B)
	\quad.
\label{additivity:rent}
\end{equation}

\item
\textit{``Pythagorean theorem''} \cite{petz:book}:
\begin{equation}
	S(\rho\|{\cal P}\sigma)=S(\rho\|{\cal P}\rho)+S({\cal P}\rho\|{\cal P}\sigma)
	\quad.
\label{pythagorean}
\end{equation}

\item
\textit{Quantum-classical interface:}
\begin{equation}
	S({\cal P}_{\{P_i\}}\rho\|{\cal P}_{\{P_i\}}\sigma)=S(\{\mbox{\rm tr}(\rho P_i)\}\|\{\mbox{\rm tr}(\sigma P_i)\})
\label{old67}
\end{equation}
with the right-hand side given by definition (\ref{2star}).

\end{enumerate}

According to relation (\ref{old75}) the ordinary entropy can always be expressed in terms of the relative entropy.
I will now show that the converse is also true.
To begin with, if $\rho$ and $\sigma$ commute then they have a joint eigenbasis, i.e., there is a set of projectors $\{P_i\}$ such that both ${\cal P}_{\{P_i\}}\rho=\rho$ and ${\cal P}_{\{P_i\}}\sigma=\sigma$.
Inserting these into Eq. (\ref{old67}) yields
\begin{equation}
	[\rho,\sigma]=0\quad\Rightarrow\quad S(\rho\|\sigma)=S(\{\mbox{\rm tr}(\rho P_i)\}\|\{\mbox{\rm tr}(\sigma P_i)\})
	\quad.
\label{5star}	
\end{equation}
It is possible to extend the Hilbert space from dimension $d$ to a larger dimension $D$, $D\geq d$, and to define in this extended Hilbert space new states 
\begin{equation}
	\tilde{\rho}:=\sum_i \frac{\mbox{\rm tr}(\rho P_i)}{\mbox{\rm tr}\tilde{P}_i} \tilde{P}_i
	\quad,\quad
	\tilde{\sigma}:=\sum_i \frac{\mbox{\rm tr}(\sigma P_i)}{\mbox{\rm tr}\tilde{P}_i} \tilde{P}_i
\end{equation}
which still commute, and where now $\sum_i\tilde{P}_i=I_D$.
Under such an extension relevant probabilities are conserved in the sense that
\begin{equation}
	\mbox{\rm tr}(\tilde{\rho}\tilde{P}_i)=\mbox{\rm tr}({\rho}{P}_i)
	\quad,\quad
	\mbox{\rm tr}(\tilde{\sigma}\tilde{P}_i)=\mbox{\rm tr}({\sigma}{P}_i)
	\quad;
\end{equation}
which in combination with Eq. (\ref{5star})---applied to both the original and the extended states---implies the conservation of relative entropy
\begin{equation}
	S(\rho\|\sigma)=S(\tilde{\rho}\|\tilde{\sigma})
	\quad.
\end{equation}
For sufficiently large (possibly infinite) $D$ the new dimensionalities $\{\mbox{\rm tr}\tilde{P}_i\}$ can be chosen such that to arbitrary precision
\begin{equation}
	\mbox{\rm tr}(\sigma P_i)=\mbox{\rm tr}\tilde{P}_i/D
\end{equation}
and thus
\begin{equation}
	\tilde{\sigma}=1/D
	\quad.
\end{equation}
For this particular choice of dimensions the relative entropy reduces to a difference of ordinary entropies:
\begin{equation}
	S(\rho\|\sigma)=S[1/D]-S[\tilde{\rho}]
	\quad.
\end{equation}
In the more general case where $\rho$ and $\sigma$ need not commute one can make use of the Pythagorean theorem (\ref{pythagorean}) with coarse-graining ${\cal P}_{\{P_i^\sigma\}}$, where the $\{P_i^\sigma\}$ project onto the eigenspaces of $\sigma$, to obtain both
\begin{equation}
	S(\rho\|\sigma)=S(\rho\|{\cal P}_{\{P_i^\sigma\}}\rho)+ S({\cal P}_{\{P_i^\sigma\}}\rho\|\sigma)
\end{equation}
and
\begin{equation}
	S(\rho\|1/d)=S(\rho\|{\cal P}_{\{P_i^\sigma\}}\rho)+ S({\cal P}_{\{P_i^\sigma\}}\rho\|1/d)
	\quad.
\end{equation}
Here we have used ${\cal P}_{\{P_i^\sigma\}}\sigma=\sigma$ and ${\cal P}_{\{P_i^\sigma\}}(1/d)=1/d$, respectively.
Combining these two results into
\begin{equation}
	S(\rho\|\sigma)=S({\cal P}_{\{P_i^\sigma\}}\rho\|\sigma)+ S(\rho\|1/d) - S({\cal P}_{\{P_i^\sigma\}}\rho\|1/d)
	\quad,
\end{equation}
the right-hand side now contains only relative entropies between states that commute.
Hence one may proceed with the same Hilbert space extension as above to express again the relative entropy in terms of ordinary entropies,
\begin{equation}
	S(\rho\|\sigma)= S[1/D]- S[\widetilde{{\cal P}_{\{P_i^\sigma\}}\rho}]+ S[{\cal P}_{\{P_i^\sigma\}}\rho]- S[\rho]
	\quad,
\label{7star}	
\end{equation}
even when $\rho$ and $\sigma$ do not commute.
So ultimately one can go full circle to reduce the problem of quantising relative entropy back to that of quantising ordinary entropy.

As for the ordinary quantum entropy, the following four properties are implied by those of the quantum relative entropy:

\begin{enumerate}

\item
\textit{Unitary invariance:} 
The invariance (\ref{groupinvariant:rent}) of relative entropy and of the uniform distribution, $g(1/d)=1/d$, together entail 
\begin{equation}
	S[g(\rho)]=S[\rho]
	\quad.
\end{equation}

\item
\textit{Invariance under Hilbert space reduction:}
If $\mbox{\rm supp}\;\rho\subseteq a$ then $P\rho P=\rho$, where $P$ projects onto $a$;
and (provided $\rho$ is normalised) the associated coarse-graining (\ref{4star}) yields ${\cal P}_{\{P,I-P\}}\rho=P/\mbox{\rm tr}P$.
With this particular coarse-graining and $\sigma=1/d$ the Pythagorean theorem (\ref{pythagorean}) reads
\begin{equation}
	S(\rho\|1/d)=S(\rho\|P/\mbox{\rm tr}P)+ S(P/\mbox{\rm tr}P\|1/d)
	\quad.
\label{1star}	
\end{equation}
By invariance (\ref{embedding:rent}) of the relative entropy under Hilbert space reduction it is
\begin{equation}
	S(\rho\|P/\mbox{\rm tr}P)=S(\rho|_a\|1/\mbox{\rm tr}P)
	\quad;
\end{equation}
so Eq. (\ref{1star}) combined with the calibration $S[\rho]=S[\rho|_a]=0$ for pure $\rho$ implies
\begin{equation}
	S(P/\mbox{\rm tr}P\|1/d)=S[1/d]-S[1/\mbox{\rm tr}P]
	\quad,
\end{equation}
and for general $\rho$
\begin{equation}
	S\left[\rho|_a\right]=S[\rho]
	\quad.
\end{equation}

\item
\textit{Additivity:}
The direct product of two pure (constituent) states is again a pure (composite) state.
Likewise the direct product of two uniform (constituent) distributions gives the uniform (composite) distribution.
The additivity (\ref{additivity:rent}) of relative entropy applied to pure $\rho$ and uniform $\sigma$ then yields
\begin{equation}
	S[1_{A\times B}/d_{A\times B}]=S[1_A/d_A]+S[1_B/d_B]
	\quad,
\end{equation}
and applied to arbitrary $\rho$
\begin{equation}
	S[\rho_A\otimes \rho_B] = S[\rho_A] + S[\rho_B]
	\quad.
\end{equation}

\item
\textit{Subadditivity:}
With the decorrelator (\ref{4startilde}) as coarse-graining and $\sigma=1/d$ the Pythagorean theorem (\ref{pythagorean}) reads
\begin{equation}
	S(\rho_{AB}\|1/d)=S(\rho_{AB}\|\rho_A\otimes \rho_B) + S(\rho_A\otimes \rho_B\|1/d)
	\quad.
\end{equation}
As the relative entropy $S(\rho_{AB}\|\rho_A\otimes \rho_B)$ is always positive, this implies
\begin{equation}
	S[\rho_{AB}]\leq S[\rho_A]+S[\rho_B]
	\quad.
\label{old82}	
\end{equation}

\end{enumerate}

\noindent
These four properties, combined with some natural assumptions regarding good mathematical behavior (continuity, extendibility to infinite dimension), determine the ordinary quantum entropy uniquely:
Up to a numerical factor it must coincide with the von Neumann entropy (\ref{old5}) \cite{ochs:axiomatic}.
Inserting this result into (\ref{7star}) then yields the quantum relative entropy (\ref{old8}), Q.E.D.

\subsection{Some properties}

For completeness I note some properties of the two quantum entropies that are not contained in, and hence are consequences of, the above axioms.
More properties can be found in Refs. \cite{araki+lieb,wehrl:rmp,thirring:book,ruskai:inequalities}.
\begin{enumerate}
\renewcommand{\theenumi}{\roman{enumi}}
\item
The relative entropy vanishes if and only if distributions are equal,
\begin{equation}
	S(\rho\|\sigma)= 0\quad\Leftrightarrow\quad \rho=\sigma
	\quad.
\end{equation}
\item
It scales with the normalisation of its arguments,
\begin{equation}
	S(\alpha\rho\|\alpha\sigma)=\alpha S(\rho\|\sigma)
	\quad\forall\ \alpha>0
	\quad;
\end{equation}
and
\item
is quasi-linear in its first argument in the sense that for $t\in[0,1]$ 
\begin{eqnarray}
	S(t\rho+(1-t)\mu\|\sigma)&=&t S(\rho\|\sigma)+ (1-t) S(\mu\|\sigma) \nonumber \\
	&& - \{S[t\rho+(1-t)\mu] - t S[\rho] -(1-t) S[\mu] \} \nonumber \\
	&& \quad
\label{old87}	
\end{eqnarray}
where the expression in $\{\cdot\}$ does \textit{not} depend on $\sigma$.
\item
Moreover, the expression in $\{\cdot\}$ is non-negative since the ordinary entropy is strictly concave,
\begin{equation}
	S[t\rho+(1-t)\mu] \geq t S[\rho] +(1-t) S[\mu]
	\quad,
\label{87star}	
\end{equation}
with equality if and only if $\rho=\mu$.
\item
The relative entropy is in general not symmetric,
\begin{equation}
	S(\rho\|\sigma)\neq S(\sigma\|\rho)
	\quad,
\end{equation}
which is plausible from its definition (\ref{3star}) via meta-probabilities:
A prior with broad support may yield evidence with narrow support, but not vice versa.
\item
However, for two infinitesimally close states with identical normalisation the relative entropy is approximately quadratic in $\delta\rho$,
\begin{equation}
	\mbox{\rm tr}(\delta\rho)=0\quad\Rightarrow\quad
	S(\rho+\delta\rho\|\rho)\sim O\left((\delta\rho)^2\right)
	\quad,
\label{7startilde}	
\end{equation}
and hence approximately symmetric.
Thus the relative entropy endows any submanifold ${\cal S}(d)|_\iota$ of states normalised to $\mbox{\rm tr}\rho=\iota\in(0,1]$ with a positive definite metric, rendering it a \textit{Riemannian manifold} \cite{balian:physrep}.
The volume element associated with this Riemannian metric will yield the proper integration measures in the de Finetti representation (\ref{definetti}) and quantum Bayes rule (\ref{old17}), provided integration is restricted to a submanifold ${\cal S}(d)|_\iota$.
\item
Again for states with identical normalisation $\iota$, any trace-preserving completely positive (TP-CP) map $\Phi$ can only decrease, but never increase their relative entropy (``monotonicity'') \cite{lindblad:monotonicity}:
\begin{equation}
	S(\Phi(\rho)\|\Phi(\sigma))\leq S(\rho\|\sigma)
	\quad\forall\ \rho,\sigma\in{\cal S}(d)|_\iota
	\quad.
\label{old182}	
\end{equation}
\item
Finally, for a composite system the ordinary quantum entropy is not only bounded from above by inequality (\ref{old82}) but also bounded from below by \cite{araki+lieb} 
\begin{equation}
	\left| S[\rho_A]-S[\rho_B]\right| \leq S[\rho_{AB}]
	\quad.
\end{equation}
This lower bound has no classical analog.
It implies in particular $S[\rho_A]=S[\rho_B]$ whenever $\rho_{AB}$ is pure.\footnote{
But it does not necessarily imply $S[\rho_A]=S[\rho_B]=0$.
}
\end{enumerate}

\section{\label{incomplete}Quantum state reconstruction with incomplete data}

In this section I turn to the second issue raised in the introduction, namely the rationale for employing the \textit{principle of minimum relative entropy} when reconstructing quantum states on the basis of incomplete data.

I begin with some notation and terminology.
The subspace $\mbox{\rm span}\{I,G_a\}$ of Liouville space shall be termed the \textit{level of description} \cite{rau:physrep}.\footnote{
Sometimes this is also called the ``observation level'' \cite{fick:book}.
}
Elements of the state manifold ${\cal S}(d)$ which satisfy the $m+1$ constraints $\langle I\rangle=\iota$ and $\{\langle G_a\rangle =g_a, a=1\ldots m\}$ form a submanifold that shall be denoted by ${\cal S}(d)|_{\iota,g}$.
That point on ${\cal S}(d)|_{\iota,g}$ which minimises the relative entropy with respect to some reference state $\sigma$ is the MinREnt distribution (\ref{old159});
it shall be denoted by $\mu^\sigma_{\iota,g}\in{\cal S}(d)|_{\iota,g}$.
There is a coarse-graining operation ${\cal P}_{I,G}^\sigma$ that maps an arbitrary state $\rho$ to 
\begin{equation}
	{\cal P}_{I,G}^\sigma\,\rho:= \mu_{\langle I\rangle_\rho,\langle G\rangle_\rho}^\sigma 
	\quad;
\end{equation}
i.e., that minimises relative entropy with respect to $\sigma$ while retaining complete information about the selected level of description.\footnote{
In the special case $\sigma=1/d$ this map is known as the \textit{Kawasaki-Gunton projector} \cite{kawasaki+gunton,fick:book,rau:physrep}.
}
This map has a number of useful properties:

\begin{enumerate}
\renewcommand{\theenumi}{\roman{enumi}}

\item
It leaves the reference state $\sigma$ unchanged,
\begin{equation}
	{\cal P}_{I,G}^\sigma\,\sigma = \sigma
	\quad.
\end{equation}

\item
It is idempotent, ${\cal P}^2={\cal P}$.
Moreover, even for $\omega\neq\sigma$ it is
\begin{equation}
	{\cal P}_{I,G}^\sigma\,{\cal P}_{I,G}^\omega = {\cal P}_{I,G}^\sigma
	\quad;
\label{idempotence}
\end{equation}
and
for arbitrary extensions $\mbox{\rm span}\{I,G_a\}\to\mbox{\rm span}\{I,G_a,F_b\}$ of the level of description 
\begin{equation}
	{\cal P}_{I,G,F}^\sigma {\cal P}_{I,G}^\sigma =
	{\cal P}_{I,G}^\sigma {\cal P}_{I,G,F}^\sigma =
	{\cal P}_{I,G}^\sigma
	\quad.
\end{equation}

\item
It satisfies the Pythagorean theorem (\ref{pythagorean}), 
\begin{equation}
		S(\rho\|{\cal P}_{I,G}^\omega\,\sigma) =S(\rho\|{\cal P}_{I,G}^\omega\,\rho)+ 
		S({\cal P}_{I,G}^\omega\,\rho\|{\cal P}_{I,G}^\omega\,\sigma)
		\quad,
\end{equation}
which for $\omega=\sigma$ simplifies to
\begin{equation}
		S(\rho\|\sigma) = S(\rho\|\mu_{\langle I\rangle_\rho,\langle G\rangle_\rho}^\sigma )+ 
		S(\mu_{\langle I\rangle_\rho,\langle G\rangle_\rho}^\sigma \|\sigma)
		\quad.
\label{old100}		
\end{equation}

\item
It encompasses the previously used coarse-grainings (\ref{4star}) and (\ref{4startilde}) as special cases.
They correspond to the choice $\omega=1/d$ and level of description $\mbox{\rm span}\{P_i\}$ or $\mbox{\rm span}\{X_A\otimes I_B,I_A\otimes X_B\}$, respectively.

\item
Finally, ${\cal P}_{I,G}^\omega$ is a trace-preserving and positive, but generally not a linear map.
In those special cases where it is linear (e.g., Eq. (\ref{4star}) but not (\ref{4startilde})) it constitutes a TP-CP map;
then by monotonicity (\ref{old182}) it reduces relative entropy:
\begin{equation}
	S(({\cal P}_{I,G}^\omega)_{\rm lin}\rho\|({\cal P}_{I,G}^\omega)_{\rm lin}\sigma)\leq S(\rho\|\sigma)
	\quad.
\end{equation}
To what extent such monotonicity also holds for general, \textit{non}-linear coarse-graining operations is an open problem worth investigating.\footnote{
This may become part of a broader effort to move beyond CP maps in the study of macroscopic dynamics \cite{shaji+sudarshan,majewski:noncpmaps}.
}

\end{enumerate}

The argument for describing macroscopic systems with MinREnt distributions of the form (\ref{old159}) now goes as follows.
In a macroscopic world the purpose of an effective description is to infer from few known averages $\iota$ and $\{g_a\}$, ascertained in $N$ trials on an exchangeable sequence, the expected values of \textit{other}---equally macroscopic---averages $\{f_b\}$ to be measured in further $M$ trials;
i.e., to determine (in vector notation)
\begin{equation}
	\vec{f}:= \int df'_1\ldots df'_l\,\mbox{\rm prob}(\Gamma^M_{f'}|\Gamma^N_{\iota,g})\,\vec{f}'
	\quad.
\end{equation}
Here, following the logic of my earlier discussion of the quantum Stein lemma (Section \ref{lemma}), I have introduced hypotheses $\Gamma$ that pertain to exchangeable sequences of length $M$ or $N$, respectively.
Yet rather than to tomographic evidence for some complete state $\rho$ these hypotheses now refer only to a small number of macroscopic averages that together do not specify a state completely.
In particular the hypothesis $\Gamma^M_{f}$ stipulates:
``After $M$ trials the measured averages for $\{F_b\}$ are within an $O(1/\sqrt{M})$-range around $\{f_b\}$'';
and likewise for $\Gamma^N_{\iota,g}$.
For large $M$ its probability of being true in a state $\sigma^{\otimes M}$---where \textit{not} necessarily $\langle F_b\rangle_\sigma=f_b$---is according to the asymptotics (\ref{quantum:stein})
\begin{equation}
	\mbox{\rm prob}(\Gamma^M_{f}|\sigma)\sim \int_{{\cal S}(d)|_{f}} d\rho\,
	\exp[-M S(\rho\|\sigma)]
	\quad,
\label{8star}	
\end{equation}
up to possibly a pre-factor that accounts for overlaps of $O(1/\sqrt{M})$-neighborhoods of the various $\rho$'s but that is independent of $\sigma$.
Inserting this likelihood function into the marginalisation
\begin{equation}
	\mbox{\rm prob}(\Gamma^M_{f'}|\Gamma^N_{\iota,g})= \int_{{\cal S}(d)} d\sigma\,
	\mbox{\rm prob}(\Gamma^M_{f'}|\sigma) \mbox{\rm prob}(\sigma|\Gamma^N_{\iota,g})
\label{14star}	
\end{equation}
and replacing
\begin{equation}
	\int df'_1\ldots df'_l \int_{{\cal S}(d)|_{f'}} d\rho\, \vec{f}'
	\quad\rightarrow\quad 
	\int_{{\cal S}(d)} d\rho\,\langle \vec{F}\rangle_\rho
\end{equation}
one obtains
\begin{equation}
	\vec{f}\sim \int_{{\cal S}(d)} d\sigma\, \mbox{\rm prob}(\sigma|\Gamma^N_{\iota,g}) 
	\int_{{\cal S}(d)} d\rho\, \exp[-M S(\rho\|\sigma)]\,\langle \vec{F}\rangle_\rho
	\quad.
\end{equation}
For sufficiently large $M$ the exponential becomes sharply peaked around $\rho=\sigma$, which (modulo normalisation) leads to
\begin{equation}
	\vec{f}\sim \int_{{\cal S}(d)} d\sigma\, \mbox{\rm prob}(\sigma|\Gamma^N_{\iota,g})\,\langle \vec{F}\rangle_\sigma
	=\mbox{\rm tr}\left(\langle\sigma\rangle_{\iota,g} \vec{F}\right)
\end{equation}
with
\begin{equation}
	\langle\sigma\rangle_{\iota,g}:= \int_{{\cal S}(d)} d\sigma\,\mbox{\rm prob}(\sigma|\Gamma^N_{\iota,g}) \cdot\sigma
	\quad.
\label{9star}	
\end{equation}
I shall argue that for large $N$ this effective single-constituent state has the MinREnt form (\ref{old159}).

The posterior $\mbox{\rm prob}(\sigma|\Gamma^N_{\iota,g})$ is related to the likelihood function $\mbox{\rm prob}(\Gamma^N_{\iota,g}|\sigma)$ via the quantum Bayes rule (\ref{metabayes}),
\begin{equation}
	\mbox{\rm prob}(\sigma|\Gamma^N_{\iota,g})= \frac{\mbox{\rm prob}(\Gamma^N_{\iota,g}|\sigma) \mbox{\rm prob}(\sigma)}{\mbox{\rm prob}(\Gamma^N_{\iota,g})}
\label{10star}	
\end{equation}
with
\begin{equation}
	\mbox{\rm prob}(\Gamma^N_{\iota,g}):= \int_{{\cal S}(d)} d\sigma'\, \mbox{\rm prob}(\Gamma^N_{\iota,g}|\sigma') \mbox{\rm prob}(\sigma')
	\quad.
\label{11star}	
\end{equation}
The likelihood function in turn is given by the integral (\ref{8star}) with replacements $M\to N$ and $f\to\iota,g$.
With the help of the Pythagorean theorem (\ref{old100}) its integrand can be factorised into
\begin{equation}
	\exp[-N S(\rho\|\sigma)]=\exp[-N S(\rho\|\mu^\sigma_{\iota,g})]\cdot \exp[-N S(\mu^\sigma_{\iota,g}\|\sigma)]
\label{8startilde}	
\end{equation}
and the second factor, which no longer depends on $\rho$, be taken out of the integral.
Applying to the argument $S(\rho\|\mu^\sigma_{\iota,g})$ of the other exponential the quadratic approximation (\ref{7startilde}) the remaining integration is over a multivariate Gaussian and---with the integration measure properly chosen to correspond to the Riemannian metric induced by the relative entropy---yields a result that is independent of $\sigma$.
Thus the $\sigma$-dependence of the likelihood function is determined entirely by the factor taken out of the integral.
For large $N$ this factor becomes sharply peaked around $\sigma=\mu^\sigma_{\iota,g}$.
One can then effectively impose $\sigma\in{\cal S}(d)|_{\iota,g}$ and restrict the integration in Eqs. (\ref{9star}) and (\ref{11star}) to ${\cal S}(d)|_{\iota,g}$, resulting finally in
\begin{equation}
	\langle\sigma\rangle_{\iota,g}\sim \int_{{\cal S}(d)|_{\iota,g}} d\sigma\, \mbox{\rm prob}(\sigma)|_{\iota,g}\, \sigma
\label{12star}
\end{equation}
with effective meta-probability on ${\cal S}(d)|_{\iota,g}$
\begin{equation}
	\mbox{\rm prob}(\sigma)|_{\iota,g}:= \frac{\mbox{\rm prob}(\sigma)}{\int_{{\cal S}(d)|_{\iota,g}} d\sigma'\, \mbox{\rm prob}(\sigma')}
	\quad\forall\ \sigma\in{\cal S}(d)|_{\iota,g}
	\quad.
\label{12startilde}	
\end{equation}

While the prior $\mbox{\rm prob}(\sigma)$ is generally not known, it is fair to assume that it is isotropic on ${\cal S}(d)$
in the sense that it is an arbitrary function of only the distance, and hence relative entropy, with respect to some (usually highly symmetric) reference state $\sigma_0$:
\begin{equation}
	\mbox{\rm prob}(\sigma) = f[S(\sigma\|\sigma_0)]
	\quad.
\end{equation}
This reference state coincides with the effective single-constituent state prior to any measurement,
\begin{equation}
	\sigma_0=\langle \sigma\rangle_{\rm prior}:= \int_{{\cal S}(d)} d\sigma\,\mbox{\rm prob}(\sigma)\,\sigma
	\quad,
\end{equation}
and hence represents any prior knowledge one may have about the system.
Applying the Pythagorean theorem (\ref{old100}) to the argument of $f$,
\begin{equation}
	S(\sigma\|\sigma_0)=S(\sigma\|\mu^{\sigma_0}_{\iota,g})+ S(\mu^{\sigma_0}_{\iota,g}\|\sigma_0)
	\quad,
\end{equation}
the posterior single-constituent state (\ref{12star}) becomes (modulo normalisation)
\begin{equation}
	\langle\sigma\rangle_{\iota,g}\sim \int_{{\cal S}(d)|_{\iota,g}} d\sigma\, 
	f\left[S(\sigma\|\mu^{\sigma_0}_{\iota,g})+ S(\mu^{\sigma_0}_{\iota,g}\|\sigma_0)\right]\cdot \sigma
	\quad.
\end{equation}
On the reduced manifold ${\cal S}(d)|_{\iota,g}$ the function $f$ is still isotropic, this time around $\mu^{\sigma_0}_{\iota,g}$;
whence indeed,
\begin{equation}
	\langle\sigma\rangle_{\iota,g}\sim \mu^{\sigma_0}_{\iota,g}
	\quad.
\label{13star}	
\end{equation}
It is this result which justifies the effective description of single constituents by means of MinREnt distributions \cite{buzek:reconstruction}.
In contrast to many textbook presentations we arrived at this result with purely statistical arguments and without any reference to chaoticity, ergodicity or other dynamical properties of the system.

One should take note, however, that in the asymptotics (\ref{12star}) the effective meta-probability $\mbox{\rm prob}(\sigma)|_{\iota,g}$ on ${\cal S}(d)|_{\iota,g}$ might still be broad, which in turn may render the effective single-constituent state insufficient to account for all macroscopic properties of interest;
in particular it is generally \textit{not} permitted to infer from the measured averages the stronger result $\sigma^{(N)}\sim (\mu^{\sigma_0}_{\iota,g})^{\otimes N}$ for the full sequence \cite{schack:bayesrule}.
This leads to the question whether a given level of description $\mbox{\rm span}\{I,G_a\}$ actually suffices to characterise a system's macrostate.
It is here that physics comes into play:
The appropriate choice for the level of description depends on the physical problem at hand and must take into account
\begin{itemize}
\item
the desired accuracy;
\item
the observables used for preparation and subsequent measurement, respectively; and
\item
in case the task is to describe a system's macroscopic dynamics, the hierarchy of time scales and hence the (possibly extended) set of ``slow'' degrees of freedom to be accounted for in a Markovian transport equation \cite{rau:physrep}.
\end{itemize}
Of course, a level of description which is so large that it encompasses the totality of observables that may ever be measured macroscopically or impact the system's macrodynamics will by definition suffice;
but such a choice will likely render the macroscopic description unduly complicated and have little predictive value.
The question is therefore whether it is feasible to \textit{contract} this maximal level of description to a smaller, manageable one.
If so, this in itself represents a non-trivial statement about the system's macroscopic properties.

The practical task amounts to checking whether the generalised canonical states (\ref{old159}) associated with a larger level of description $\mbox{\rm span}\{I,G_a,F_b\}$ and with the contracted level of description $\mbox{\rm span}\{I,G_a\}$, respectively, coincide within some prescribed error margin.
As both levels of description contain the unit operator $I$, both states will lie on the same reduced state manifold ${\cal S}(d)|_\iota$.
By Eq. (\ref{7startilde}) this manifold is endowed with a Riemannian metric, so the distance between two states is given approximately by their relative entropy.
Using the Pythagorean theorem (\ref{old100}) with $\rho=\mu^\sigma_{\iota,g,f}$ this distance can be expressed as
\begin{equation}
	S(\mu^\sigma_{\iota,g,f}\|\mu^\sigma_{\iota,g})= S(\mu^\sigma_{\iota,g,f}\|\sigma) - S(\mu^\sigma_{\iota,g}\|\sigma)
	\quad,
\end{equation}
which in the special case $\sigma=1/d$ reduces further to the difference of ordinary entropies
$(S[\mu^{1/d}_{\iota,g}]-S[\mu^{1/d}_{\iota,g,f}])$.
Thus the feasibility of a level contraction can be assessed simply by comparing entropies associated with the original and contracted levels of description, respectively.
Whether or not an entropy differential must be considered significant---and hence level contractions must stop---depends both on the number $N$ of trials and on the accuracy of the measurements performed.
By Eq. (\ref{old29}) an entropy differential is not significant as long as it is within the $O(1/N)$-range of statistical fluctuations.
Yet even when the entropy differential is outside this range it may still be considered insignificant provided
\begin{equation}
	S(\mu^\sigma_{\iota,g,f}\|\sigma) - S(\mu^\sigma_{\iota,g}\|\sigma) < S(\mu^\sigma_{\iota,g,f}\|\mu^\sigma_{\iota,g,f+\Delta f})
	\quad,
\label{15startilde}	
\end{equation}
where $\Delta f$ is the finite accuracy with which the (presumably redundant) averages $\{f_b\}$ can be measured.
The right-hand side is approximately quadratic in $\Delta f$;
an explicit formula will be given in Appendix \ref{statmech} (Eq. (\ref{secondorder}).

Successive contractions eventually lead to a smallest possible level of description which cannot be contracted further without a significant increase in entropy.
It is this smallest possible level of description which best captures the essential features of, and hence furnishes the most suitable theoretical model for, a system's macrostate.\footnote{
A lucid illustration of such an iterative diagnosis can be found in Jaynes' analysis of Wolf's die data \cite{jaynes:concentration} as recounted in my earlier didactical note \cite{rau:nutshell}.
There the iteration proceeds in the opposite direction, starting from a minimal level of description and successively enlarging it.
}

\newpage

\appendix

\section{\label{statmech}Statistical mechanics}

In this appendix I provide some link between the above conceptual discussions and the conventional formulation of statistical mechanics.

Once the appropriate level of description is established, the macrostate of a physical system can be characterised completely by the $m+1$ expectation values $\{\iota,g_a\}$ of the relevant observables.
The associated MinREnt state (\ref{old159}) has an (ordinary) entropy
\begin{equation}
	S[\mu_{\iota,g}^\sigma/\iota]+\left( \langle\ln\sigma\rangle_{\mu_{\iota,g}^\sigma/\iota} - \langle\ln\sigma\rangle_{1/d}\right) = \ln Z +\sum_{a=1}^m \lambda^a g_a/\iota
	\quad,
\label{minrent:entropy2}
\end{equation}
which in the special case $\sigma=1/d$ and $\iota=1$ reduces to the familiar relation
\begin{equation}
	S[\mu_{1,g}^{1/d}] = \ln Z +\sum_{a=1}^m \lambda^a g_a
	\quad.
\end{equation}
As
\begin{equation}
	\frac{\partial}{\partial\lambda^a}\ln Z = -g_a/\iota 
	\quad,
\label{partial:z}
\end{equation}
infinitesimal variation yields
\begin{equation}
	d \left\{S[\mu_{\iota,g}^\sigma/\iota]+\left( \langle\ln\sigma\rangle_{\mu_{\iota,g}^\sigma/\iota} - \langle\ln\sigma\rangle_{1/d}\right)\right\} =
	\sum_{a=1}^m \lambda^a d\{g_a/\iota\}
\end{equation}
with the familiar special case 
\begin{equation}
		d S[\mu_{1,g}^{1/d}] = \sum_{a=1}^m \lambda^a dg_a
		\quad.
\end{equation}

Upon infinitesimal variation of Lagrange parameters or of the state's normalisation arbitrary expectation values change according to
\begin{equation}
	d\langle A\rangle_{\mu_{\iota,g}^\sigma}= (\langle A\rangle_{\mu_{\iota,g}^\sigma}/\iota) d\iota - 
	\sum_{a=1}^m \langle \delta G_a;A\rangle_{\mu_{\iota,g}^\sigma} d\lambda^a
	\quad,
\end{equation}
where
\begin{equation}
	\delta G_a:= G_a- (g_a/\iota) I
\end{equation}
and $\langle;\rangle_\rho$ is the \textit{canonical correlation function} with respect to the state $\rho$
\begin{equation}
\langle B;A\rangle_\rho :=
\int_0^1{d}\nu\,\mbox{tr}
\left[{\rho}^\nu {B}^\dagger {\rho}^{1-\nu} {A}\right]
\quad.
\end{equation}
Considering the special case $A=G_b$ and $d\iota=0$ yields
\begin{equation}
	d g_b= -\sum_{a=1}^m d\lambda^a C_{ab}
\end{equation}
with \textit{correlation matrix}
\begin{equation}
	C_{ab}:=\langle \delta G_a;\delta G_b\rangle_{\mu^\sigma_{\iota,g}} =
	- \left(\frac{\partial g_b}{\partial\lambda^a}\right)_{\lambda,\iota}
	\quad.
\end{equation}
Since the canonical correlation function constitutes a scalar product, this correlation matrix is symmetric and positive.

The relative entropy of two different macrostates on the same level of description reads
\begin{equation}
	S(\mu_{\iota,g}^\sigma\|\mu_{\tilde{\iota},\tilde{g}}^\sigma)= \sum_{a=1}^m (\tilde{\lambda}^a-\lambda^a) g_a
	+ \iota (\ln\tilde{Z}-\ln Z) - \iota\ln(\tilde{\iota}/\iota)
	\quad.
\label{relent:neighboring}
\end{equation}
For neighboring states the right-hand side can be expanded in powers of $\Delta g_a:=\tilde{g}_a-g_a$ and $\Delta\iota:=\tilde{\iota}-\iota$.
Assuming identical normalisation ($\Delta\iota=0$) and using Eq. (\ref{partial:z}) as well as
\begin{equation}
	\iota \frac{\partial^2}{\partial\lambda^a\partial\lambda^b}\ln Z=C_{ab}
	\quad,
\end{equation}
the first non-vanishing term is of second order:
\begin{equation}
	S(\mu_{\iota,g}^\sigma\|\mu_{{\iota},{g+\Delta g}}^\sigma) \approx
	\frac{1}{2} \sum_{a,b=1}^m C_{ab} \Delta\lambda^a \Delta\lambda^b \approx
	\frac{1}{2} \sum_{a,b=1}^m (C^{-1})^{ab} \Delta g_a \Delta g_b
	\geq 0
	\ .
\label{secondorder}	
\end{equation}
If calculated on the extended level of description $\mbox{\rm span}\{I,G_a,F_b\}$, this result provides the right-hand side of Eq. (\ref{15startilde}).
To lowest order in $\Delta g$ the correlation matrix $C$ may be evaluated in either of the two neighboring states.

How the above general results lead to familiar thermodynamic relations is discussed in a separate didactical note \cite{rau:nutshell}.

\section{Emergence of classicality}

In this appendix I discuss briefly the fact that despite the quantum nature of the underlying constituents reasoning at the macroscopic level is approximately classical, in the following sense.
Measuring macroscopic averages in the fashion described in Section \ref{incomplete} constitutes a learning process that can be encapsulated in a macroscopic version of the quantum Bayes rule (\ref{metabayes}),
\begin{equation}
	\mbox{\rm prob}(\Gamma^M_f|\Gamma^N_{\iota,g})= \frac{\mbox{\rm prob}(\Gamma^N_{\iota,g}|\Gamma^M_f) \cdot\mbox{\rm prob}(\Gamma^M_f)}{\mbox{\rm prob}(\Gamma^N_{\iota,g})}
	\quad;
\label{16star}	
\end{equation}
it follows in a straightforward manner from the marginalisation (\ref{14star}) and application of the quantum Bayes rule (\ref{metabayes}) to both factors in the integrand.
When the numbers $M,N$ of trials become large then there is the even stronger result
\begin{equation}
	\mbox{\rm prob}(\Gamma^M_f|\Gamma^N_{\iota,g}) \cdot\mbox{\rm prob}(\Gamma^N_{\iota,g}) \sim \mbox{\rm prob}(\Gamma^{M+N}_{\iota,g,f})
	\quad.
\label{17star}	
\end{equation}
This has the form of the classical product rule
\begin{equation}
	\mbox{\rm prob}(a|b)\cdot\mbox{\rm prob}(b)=\mbox{\rm prob}(a\cap b)
\end{equation}
and thus indicates the emergence of classical probability calculus in the macroscopic limit.

The proof of this asymptotic product rule proceeds as follows.
In the marginalisation (\ref{14star}) one can apply the quantum Bayes rule (\ref{metabayes}) to the second factor in the integrand to obtain
\begin{equation}
	\mbox{\rm prob}(\Gamma^M_f|\Gamma^N_{\iota,g}) \mbox{\rm prob}(\Gamma^N_{\iota,g}) = 
	\int_{{\cal S}(d)} d\sigma\,\mbox{\rm prob}(\Gamma^M_f|\sigma) \mbox{\rm prob}(\Gamma^N_{\iota,g}|\sigma) 
	\mbox{\rm prob}(\sigma)
	\quad.
\label{18star}	  
\end{equation}
The first two factors on the right-hand side are given asymptotically by Eq. (\ref{8star}), and the ensuing product of exponentials can be re-expressed by quasi-linearity (\ref{old87}) to give (modulo normalisation)
\begin{eqnarray}
	\mbox{\rm prob}(\Gamma^M_f|\sigma) \mbox{\rm prob}(\Gamma^N_{\iota,g}|\sigma) &\sim&
	\int_{{\cal S}(d)|_f} d\rho \int_{{\cal S}(d)|_{\iota,g}} d\omega\, D^{M,N}_{\rho,\omega} \times \nonumber \\
	&& \times \exp\left[-(M+N)\,S\left(\left.\frac{M}{M+N}\rho + \frac{N}{M+N}\omega\right\|\sigma\right)\right] \nonumber \\
	&& \quad
\end{eqnarray}
with
\begin{eqnarray}
	D^{M,N}_{\rho,\omega}&:=& \exp\left[-(M+N) \left\{ S\left[ \frac{M}{M+N}\rho + \frac{N}{M+N}\omega \right]\right.\right.
	 \nonumber \\
	&& \quad\quad\quad\quad\quad \left.\left. - \frac{M}{M+N} S[\rho] - \frac{N}{M+N} S[\omega] \right\}\right]
	\quad.
\end{eqnarray}
By strict concavity (\ref{87star}) this newly defined function is for large $M,N$ sharply peaked around $\rho=\omega$.
The latter equality is possible only if $\rho$ lies on the reduced manifold 
${\cal S}(d)|_{\iota,g,f}={\cal S}(d)|_{f}\cap {\cal S}(d)|_{\iota,g}$, which I assume to be non-empty.
Then one may replace in the asymptotic limit (modulo normalisation)
\begin{equation}
	\int_{{\cal S}(d)|_f} d\rho \int_{{\cal S}(d)|_{\iota,g}} d\omega\, D^{M,N}_{\rho,\omega} 
	\quad\rightarrow\quad
	\int_{{\cal S}(d)|_{\iota,g,f}} d\rho
\end{equation}
and
\begin{equation}
	S\left(\left.\frac{M}{M+N}\rho + \frac{N}{M+N}\omega\right\|\sigma\right)
	\quad\rightarrow\quad
	S(\rho\|\sigma)
	\quad;
\end{equation}
whence
\begin{equation}
	\mbox{\rm prob}(\Gamma^M_f|\sigma) \cdot\mbox{\rm prob}(\Gamma^N_{\iota,g}|\sigma) \sim
	\mbox{\rm prob}(\Gamma^{M+N}_{\iota,g,f}|\sigma)
	\quad,
\end{equation}
which in turn by Eqs. (\ref{11star}) and (\ref{18star}) implies the asymptotic product rule (\ref{17star}), Q.E.D.

\newpage

\newcommand{\etalchar}[1]{$^{#1}$}

\end{document}